\documentclass[conference]{IEEEtran}
\IEEEoverridecommandlockouts

\usepackage{cite}
\usepackage{amsmath,amssymb,amsfonts}
\usepackage{algorithmic}
\usepackage{graphicx}
\usepackage{textcomp}
\usepackage{xcolor}
\usepackage{braket}
\usepackage{subcaption}

\def\BibTeX{{\rm B\kern-.05em{\sc i\kern-.025em b}\kern-.08em
    T\kern-.1667em\lower.7ex\hbox{E}\kern-.125emX}}
\begin{document}

\bstctlcite{IEEEexample:BSTcontrol}

\title{Quantum Arithmetic Algorithms: Implementation, Resource Estimation, and Comparison}

\author{\IEEEauthorblockN{Dmytro Fedoriaka}
\IEEEauthorblockA{\textit{University of Washington}\\
Seattle, Washington, USA\\
fedimser@cs.washington.edu}
\and
\IEEEauthorblockN{Brian Goldsmith}
\IEEEauthorblockA{\textit{Independent Researcher}\\
Buena Park, California, USA\\
brian.goldsmith@gmail.com}
\and
\IEEEauthorblockN{Yingrong Chen}
\IEEEauthorblockA{\textit{Microsoft Quantum}\\
Redmond, Washington, USA\\
yingrongchen@microsoft.com}
}

\maketitle

\begin{abstract}
As quantum computing technology advances, the need for optimized arithmetic circuits continues to grow. This paper presents the implementation and resource estimation of a library of quantum arithmetic algorithms, including addition, multiplication, division, and modular exponentiation. Using the Azure Quantum Resource Estimator, we evaluate runtime, qubit usage, and space-time trade-offs and identify the best-performing algorithm for each arithmetic operation. We explore the design space for division, optimize windowed modular exponentiation, and identify the tipping point between multipliers, demonstrating effective applications of resource estimation in quantum research. Additionally, we highlight the impact of parallelization, reset operations, and uncomputation techniques on implementation and resource estimation. Our findings provide both a practical library and a valuable knowledge base for selecting and optimizing quantum arithmetic algorithms in real-world applications. 
\end{abstract}


\section{Introduction}

Quantum algorithms could revolutionize computing by solving problems that are infeasible for classical systems; however, their practical implementation remains challenging because of the gap between theoretical requirements and current quantum hardware. Before fully realizing quantum computers, systematic resource estimation is essential for evaluating algorithmic designs across different quantum machine models and error correction schemes. Previous studies have estimated the resources required for end-to-end applications such as Trotter-based dynamic simulations, double-factorized qubitization in quantum chemistry, and Shor's algorithm \cite{beverland2022}, as well as specific quantum primitives such as quantum multiplication \cite{hansen2023}. These studies provide critical insights into the feasibility and scalability of different quantum applications as hardware continues to evolve.

Quantum arithmetic algorithms are essential components of advanced quantum computing, and improving their efficiency can significantly reduce resource requirements of real-world quantum applications\cite{ArithmeticReview2024}. In this work, we implemented a comprehensive suite of quantum arithmetic algorithms for many operations---including addition, multiplication, division, and modular exponentiation---and evaluated their performance using the Azure Quantum Resource Estimator \cite{Azure_Quantum_Resource_Estimator}. This tool provides detailed estimates of the logical and physical resources, as well as the runtime, required to execute a circuit on a fault-tolerant quantum computer. We identified the best algorithms for each arithmetic operation based on qubit count and runtime. Additionally, we explored different algorithmic building blocks, optimized parameters, analyzed asymptotic complexities, and identified ``tipping points" where specific algorithms become advantageous. We outlined practical strategies for quantum researchers to seamlessly integrate the Azure Quantum Resource Estimator into their workflows.

The remainder of the paper is organized as follows. Section \ref{sec:AlgorithmsOverview} introduces the implemented algorithms. Section \ref{sec:Methods} describes the experimental setup. Section \ref{sec:ResultsDiscussion} presents the results, explores different applications of resource estimation, and offers insights into quantum arithmetic algorithm design.

\section{Algorithms Overview} \label{sec:AlgorithmsOverview}
This section summarizes the implemented algorithms for six arithmetic operations. We assigned each implementation a short code name for presenting the experimental results. All circuits implement a unitary matrix and act on unsigned quantum integers.

\subsection{Addition}

\textbf{In-place addition}. An in-place adder $U$ acts on two $n$-qubit registers as:  $$U \ket{a} \ket{b} = \ket{a} \ket{(a+b) \% 2^n}$$ Here ``\%" denotes the modulo operation. Addition modulo $2^n$ is used for convenience, and it can be converted to a non-modular adder by adding one output qubit for the carry, and vice versa\cite{cuccaro2004new}. Two common approaches for addition are Ripple-Carry-Adder (RCA) and Carry-Lookahead Adder (CLA). We considered the following in-place adders:
\begin{itemize}
  \item Gidney --- RCA by Craig Gidney \cite{gidney2018halving}.
  \item TTK --- RCA by Takahashi et al. \cite{takahashi2009quantum}.
  \item CDKM --- RCA by Cuccaro et al. \cite{cuccaro2004new}.
  \item DKRS --- CLA by Draper et al. \cite{draper2006logarithmic}.
  \item JHHA --- adder by Jayashree et al. \cite{jayashree2016ancilla}.
  \item TR --- RCA by Thapliyal and Ranganathan \cite{thapliyal2013design}.
  \item QFT --- adder using Quantum Fourier Transform (QFT) \cite{draper2000}.
\end{itemize}

For Gidney, TTK and QFT in-place adders, we used implementations from the Q\# library(\texttt{RippleCarryCGIncByLE}, \texttt{RippleCarryTTKIncByLE}, \texttt{FourierTDIncByLE})\cite{qsharpStdLib}.

\textbf{Out-of-place addition}. An out-of-place adder $U$ acts on three $n$-qubit registers as: $$U \ket{a} \ket{b} \ket{0} = \ket{a}  \ket{b} \ket{(a+b) \% 2^n}.$$ We considered the following out-of-place adders:

\begin{itemize}
    \item Gidney --- RCA by Craig Gidney \cite{gidney2018halving}.
    \item DKRS --- CLA by Draper et al. \cite{draper2006logarithmic}.
    \item CT --- RCA by Cheng and Tseng \cite{cheng2002quantum}.
    \item LingStruct --- adder using Ling Structure \cite{LingAdder2023}.
    \item Gayathri --- RCA by Gayathri et al. \cite{gayathri2021}.
    \item Wang --- RCA by Wang et al. \cite{wang2016improved}.
    \item HiRadix --- higher radix adder from \cite{wang2023higher}, using radix $r=\lceil n/2 \rceil$.
\end{itemize}

For Gidney and DKRS out-of-place adders, we used implementations from the Q\# library (\texttt{RippleCarryCGAddLE}, \texttt{LookAheadDKRSAddLE}) \cite{qsharpStdLib} .

\textbf{Quantum-classical addition.} A quantum-classical adder $U_a$, given a classical integer constant $a \in [0, 2^n)$, acts on an $n$-qubit register as: $$U_a \ket{b} = \ket{(a+b) \% 2^n}.$$ A straightforward implementation uses an in-place adder with $n$ ancillas \cite{liu2021cnot}, available in the Q\# library \cite{qsharpStdLib} (\texttt{IncByLUsingIncByLE}). We implemented the following quantum-classical adders:
\begin{itemize}
    \item TTK, Gidney, JHHA, CDKM, DKRS --- quantum-classical adders built from in-place adders using \texttt{IncByLUsingIncByLE}.
    \item ConstAdder --- quantum-classical adder proposed in \cite{fedoriaka2025}.
    \item QFT --- QFT-based quantum-classical adder \cite{draper2000,pavlidis2014fast}.
\end{itemize}

\subsection{Multiplication}

A multiplier $U$ acts on three registers of $n$, $n$ and $2n$ qubits as: $$U \ket{a} \ket{b} \ket{0} = \ket{a} \ket{b} \ket{a \cdot b}.$$ We implemented the following multipliers:
\begin{itemize}
  \item JHHA --- Shift-and-Add algorithm from \cite{jayashree2016ancilla}.
  \item MCT --- Shift-and-Add algorithm from \cite{munozcoreas2017}.
  \item Schoolbook --- ``Schoolbook" algorithm from \cite{gidneyKaratsubaRepo}.
  \item Karatsuba --- Karatsuba algorithm proposed in \cite{gidney2019karatsuba}, implemented in \cite{gidneyKaratsubaRepo}.
  \item Karatsuba-8 --- same as above, but using minimal ``piece size" of 8 (as opposed to 32 as proposed in \cite{gidney2019karatsuba}).
  \item Wallace Tree --- multiplier using the Wallace Tree proposed in \cite{orts2023improving}.
\end{itemize}

\subsection{Division}

A divider $U$ acts on three $n$-qubit registers as: $$U \ket{a} \ket{b} \ket{c} = \ket{a\%b} \ket{b} \ket{a / c}.$$  We implemented the following dividers:
\begin{itemize}
  \item Restoring and non-restoring (denoted as ``R" and ``NR") --- as described in \cite{thapliyal2019quantum}.
  \item AKBF --- a restoring division as described in \cite{khosropour2011quantum}.
\end{itemize}

All these dividers use in-place adders as building blocks. Since their original implementation used earlier adders from \cite{draper2000} and \cite{Thapliyal2016MappingOS}, we optimize them by exploring alternative adder designs \cite{wang2024boosting}. We implemented divider circuits that take the adder as a parameter and evaluated all combinations of the two divider types (restoring and non-restoring) with four in-place adders: Gidney \cite{gidney2018halving}, TTK \cite{takahashi2009quantum}, CDKM \cite{cuccaro2004new}, JHHA \cite{jayashree2016ancilla}. We omitted results for DKRS and QFT in-place adders as they exhibited significantly worse metrics. For the subtractor block within the divider, we adapted the in-place adders to compute $\overline{\bar{a} + b}$ ($\overline{x}$ being the binary complement of $x$), which is equivalent to $ a - b $ \cite{Thapliyal2016MappingOS}. 

\subsection{Modular exponentiation (ModExp)}

Given classical numbers $a$ and $N$, ModExp circuit $U_{a,N}$ acts on two $n$-qubit registers as follows: $$U_{a,N}\ket{x} \ket{0} = U_{a,N}\ket{x} \ket{a^x \% N}.$$ This operation is an important building block for Shor's algorithm \cite{shor1994algorithms}. 

We implemented the following ModExp algorithms:
\begin{itemize}
  \item LYY --- ModExp from \cite{liu2021cnot}, without windowing or intermediate data accumulation.
  \item LYY-W-1, LYY-W-11 --- ModExp from \cite{liu2021cnot} using fast modular multiplication and windowing, with window sizes of 1 and 11.
  \item LYY-W-Opt --- ModExp from \cite{liu2021cnot} using near-optimal window size $w = \lfloor 2 \log_2(n) + 0.5 \rfloor$ (see details in \S \ref{parameter_tuning}).
  \item LYY-MW-1, LYY-MW-11 --- ModExp from \cite{liu2021cnot} using Montgomery modular multiplication and windowing, with window sizes of 1 and 11.
  \item QFT --- QFT-based ModExp based on Granlund-Montgomery division \cite{pavlidis2014fast}.
\end{itemize}

\subsection{Other arithmetic operations}
We also implemented other arithmetic operations, including incrementer \cite{li2014class}, subtraction\cite{TR2009Sub}\cite{Thapliyal2016MappingOS}, comparator \cite{Xia2019NovelMQ}, table lookup \cite{babbush2018encoding}, square root \cite{munoz2018squareRoot}, and greatest common divisor \cite{saeedi2013}. Due to space limitations, these algorithms are not described in detail here.

\section{Methods} \label{sec:Methods}
\subsection{Implementation}

We implemented the algorithms described above in Q\# \cite{Svore_2018}, as a Q\# external library. The implementation is available on GitHub (https://github.com/fedimser/quant-arith-re). These algorithms are designed and thoroughly tested to work with registers of any size $n$. The input is always encoded as a little-endian \texttt{Qubit} array, where the least significant bit is stored in the first qubit. All numbers are treated as unsigned integers.

\subsection{Resource Estimation}

For the algorithms described above, we ran resource estimation using the Azure Quantum Resource Estimator available as part of Azure Quantum Development Kit \cite{Azure_Quantum_Resource_Estimator}. We used the default parameters \cite{Lopez_2024}, including a gate-based instruction set with operation times and fidelities corresponding to superconducting transmon qubits, a surface code quantum error correction scheme, and an overall tolerated error budget of 0.001.

Many quantum applications, including Shor’s algorithm, require circuits whose complexity grows with n, so we generated circuits based on an input size $n$. The estimations were carried out on a logarithmic grid to explore a range of computationally feasible problem sizes, starting at $n=3$ with a step size of $2^{1/4}$. We tested larger values of $n$ until the execution time exceeded several hours. Below are experimental details for each class of algorithms:
\begin{itemize}
    \item \textbf{In-place addition}. $n$ is the size of two registers. We ran experiments  up to $n=2^{20}$.
    \item \textbf{Out-of-place addition}. $n$ is the size of each of three registers. We ran experiments up to $n=2^{20}$.
    \item \textbf{Quantum-classical addition}. $n$ is the size of the single register. We ran experiments up to $n=2^{20}$. The constant value $a=\sum_{i=0}^{\lceil n/2 \rceil} 4^i$ was used.
    \item \textbf{Multiplication}. $n$ is the size of the input registers and the output register has $2n$ qubits. We ran experiments up to $n=2^{15}$.
    \item \textbf{Division}. $n$ is the size of each of the three registers. We ran experiments up to $n=2^{14}$.
    \item \textbf{Modular exponentiation}. $n$ is the size of both registers. We ran experiments up to $n=2^{9}$. We used constant values of $N=2^n-1$ and $a=5^{24}+24^5 \approx 5.9\cdot 10^{16}$.
\end{itemize}

We compared resource estimation metrics across different algorithms within the same class. This is a fair comparison, since all algorithms within a class solve the same problem. For some resource-heavy algorithms, experiments were stopped before reaching maximum $n$. Our selection of $n$ aligns with problem sizes that are relevant for near-term quantum applications. 

\subsection{Pareto Frontier Estimation}

The Pareto frontier estimation in Azure Quantum Resource Estimator explores the trade-offs between qubit count and runtime by running the circuit repetitively with different number of T-factories available. The error budget, error rates, and error correction properties, such as code distance, are used to determine the boundaries and steps to adjust the number of T-factories. By reducing the number of T-factories, the qubit count decreases but runtime increases as the circuit may need to wait for T-state availability. The Pareto frontier represents the set of optimal configurations for T-factory parallelism. Multiple estimates are taken with varying T-factory availability, so the largest possible input size ($n$) is chosen to observe pronounced trade-offs while managing computational resources. We set up two experiments at a fixed input size:

\begin{itemize}
    \item \textbf{Addition}. This experiment included in-place, out-of-place, and quantum-classical adders. The input register size was set to $n=2^{11}$ as QFT-based algorithms could not produce results beyond this due to resource constraints.
    \item \textbf{Multiplication/Division}. This experiment included multipliers and dividers. The input register size was set to $n=2^{9}$ as some restoring division algorithms could not produce results beyond this due to resource constraints..
\end{itemize}

To better visualize the results, we combined algorithms with resource estimates within 1\% of each other. In the adder experiment, the Gidney* result includes Gidney\cite{gidney2018halving}, Gayathri\cite{gayathri2021}, and Wang\cite{wang2016improved} adders. The TTK† result combines TTK\cite{takahashi2009quantum}, CDKM\cite{cuccaro2004new}, and JHHA\cite{jayashree2016ancilla} adders. The DKRS§ result represents DKRS\cite{draper2006logarithmic}, HiRadix\cite{wang2023higher}, and CT\cite{cheng2002quantum} adders. For multipliers and dividers, the NR+TTK‡ result combines NR+TTK, NR+CDKM, and NR+JHHA dividers.

\begin{figure*}
    \centering
    \noindent\makebox[\textwidth]{
      \includegraphics[width=0.53\textwidth]{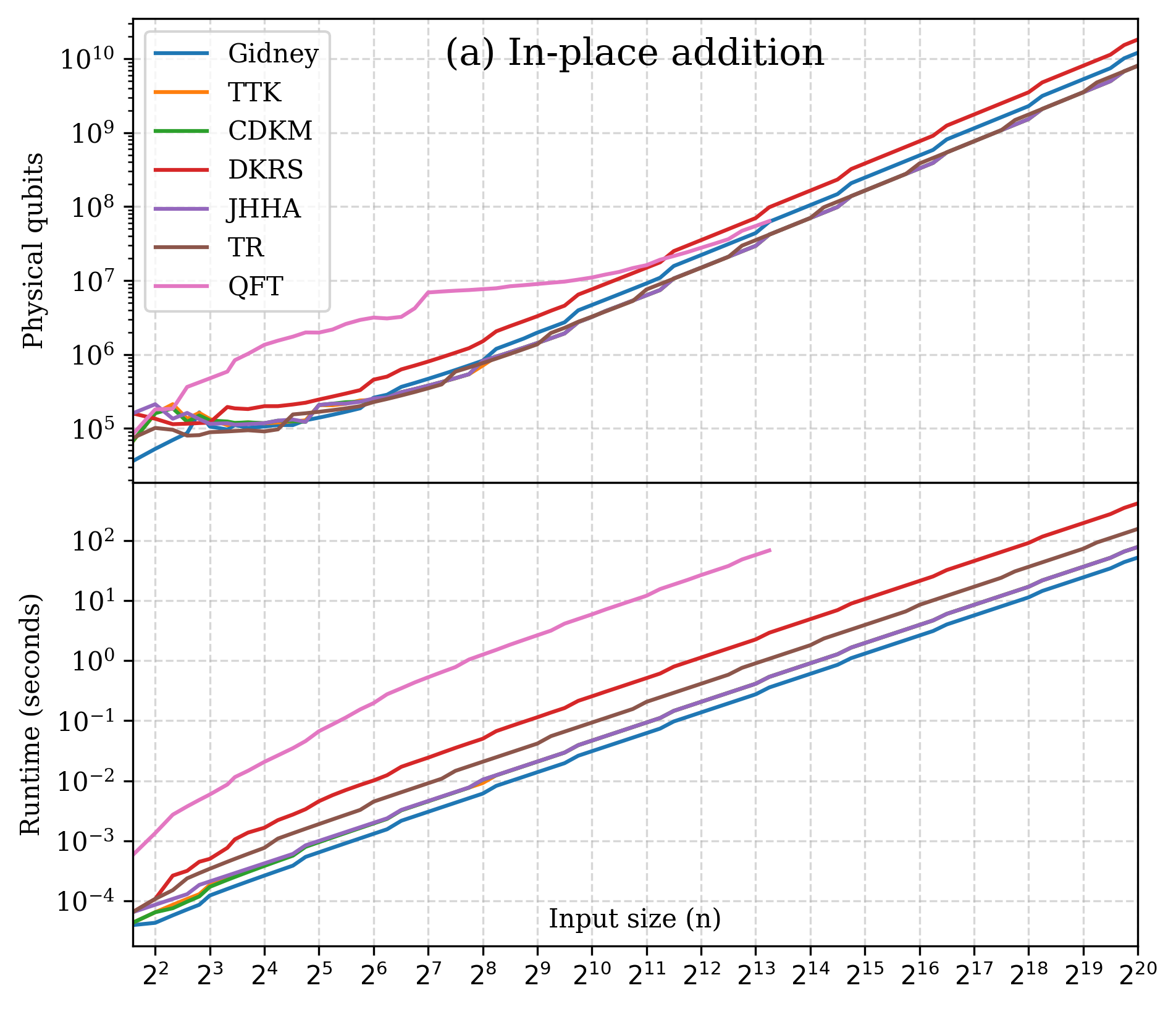}
      \includegraphics[width=0.53\textwidth]{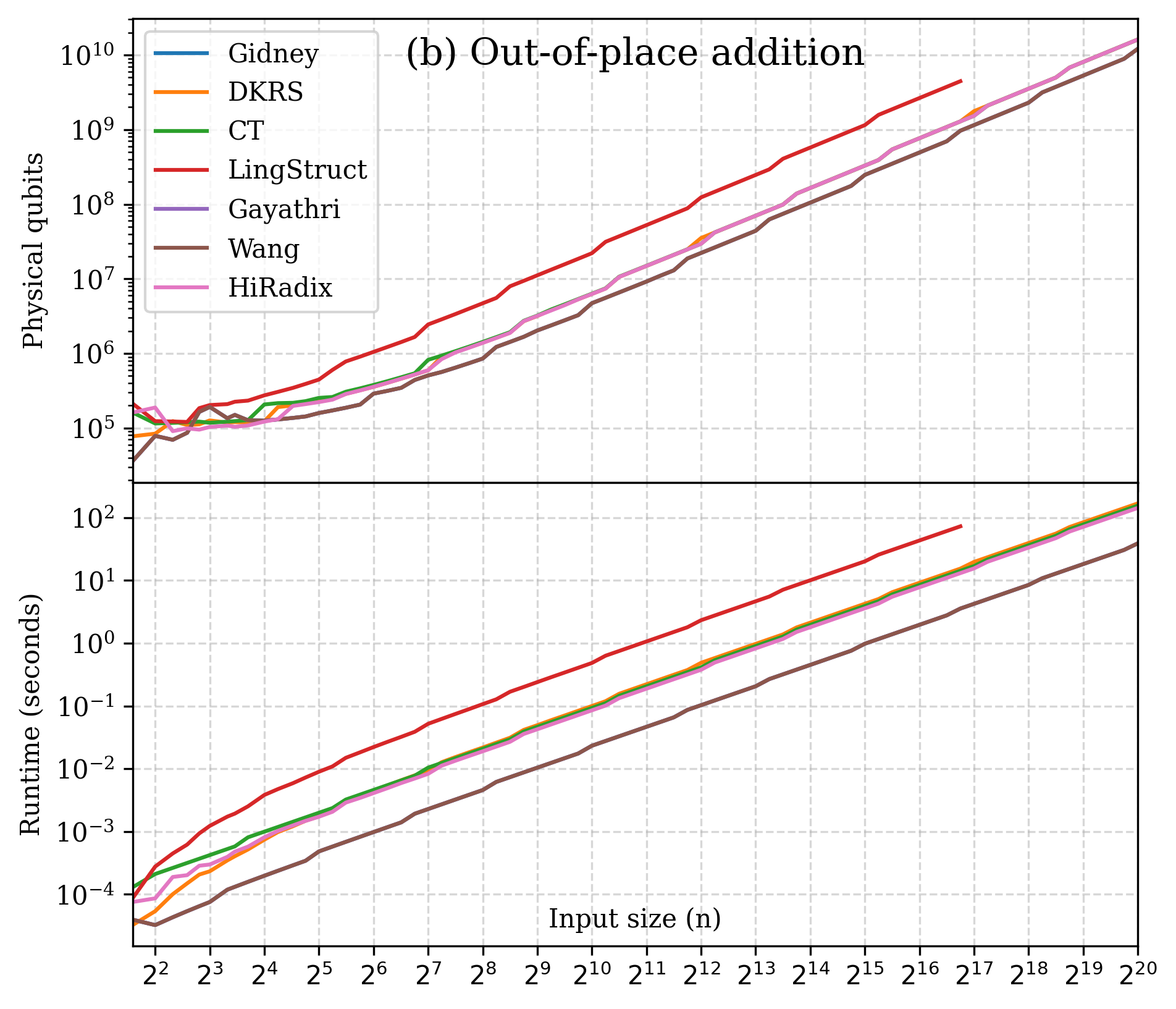}
    }
    \noindent\makebox[\textwidth]{
      \includegraphics[width=0.53\textwidth]{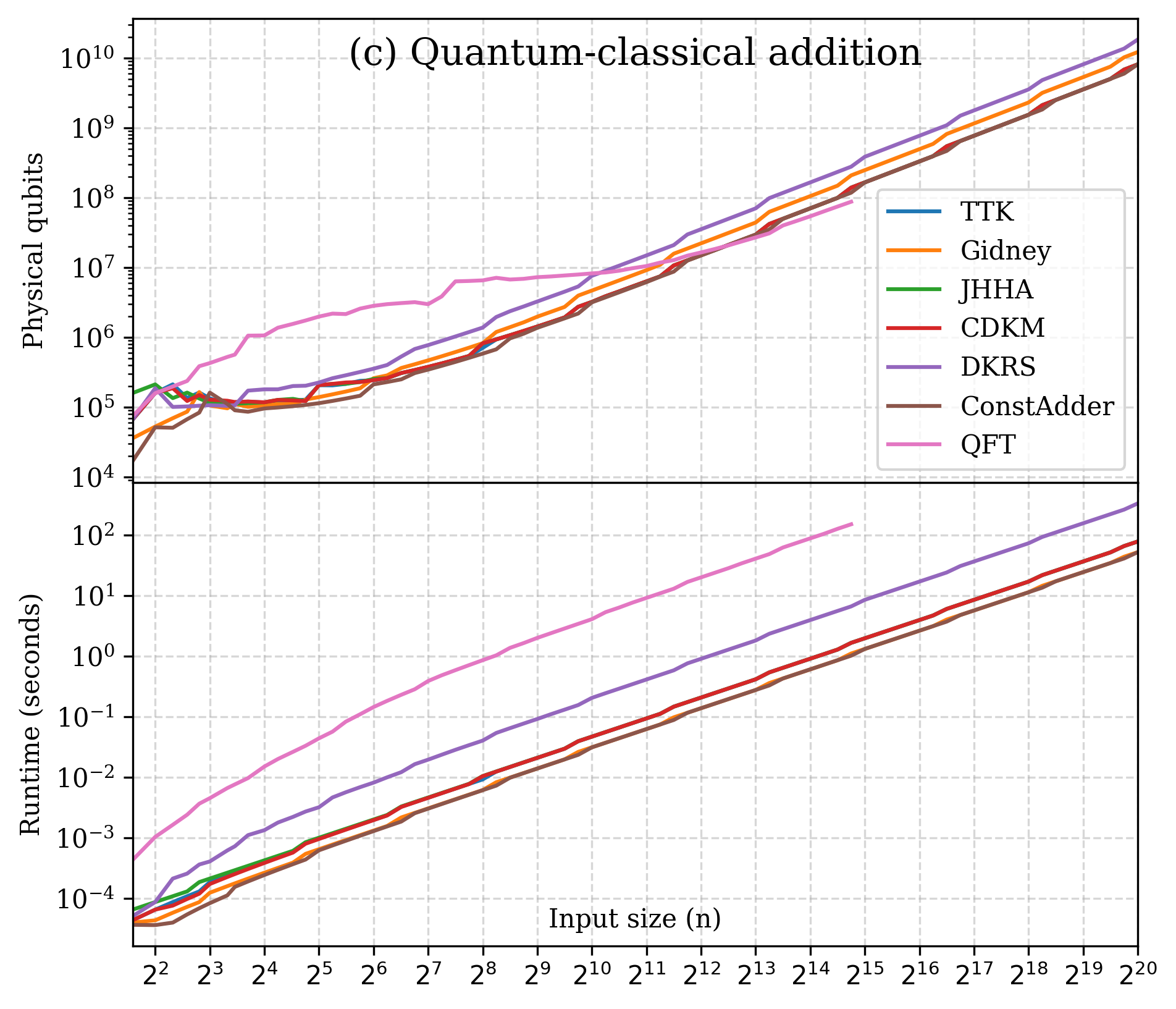}
      \includegraphics[width=0.53\textwidth]{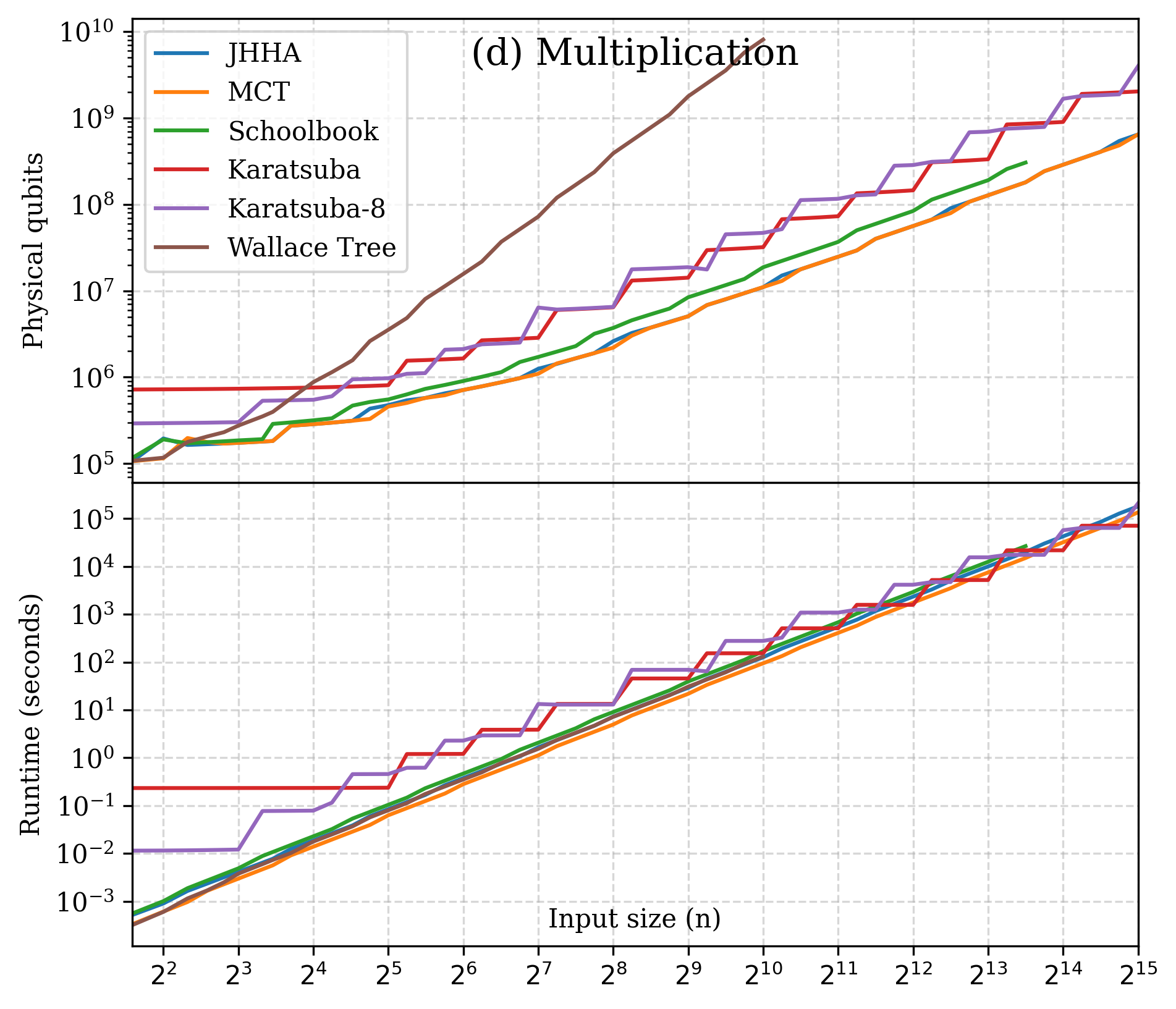}
    }
    \noindent\makebox[\textwidth]{
      \includegraphics[width=0.53\textwidth]{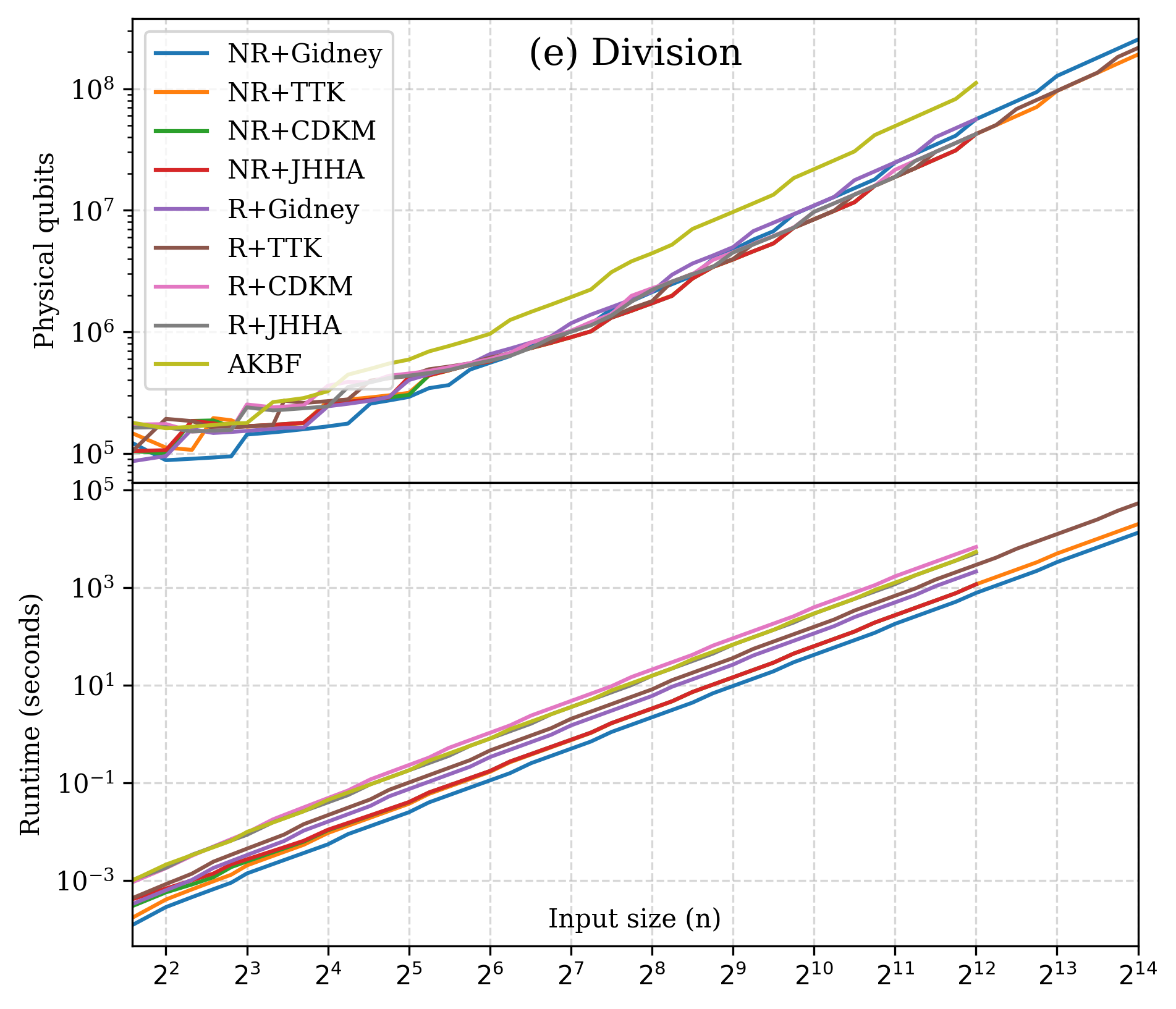}
      \includegraphics[width=0.53\textwidth]{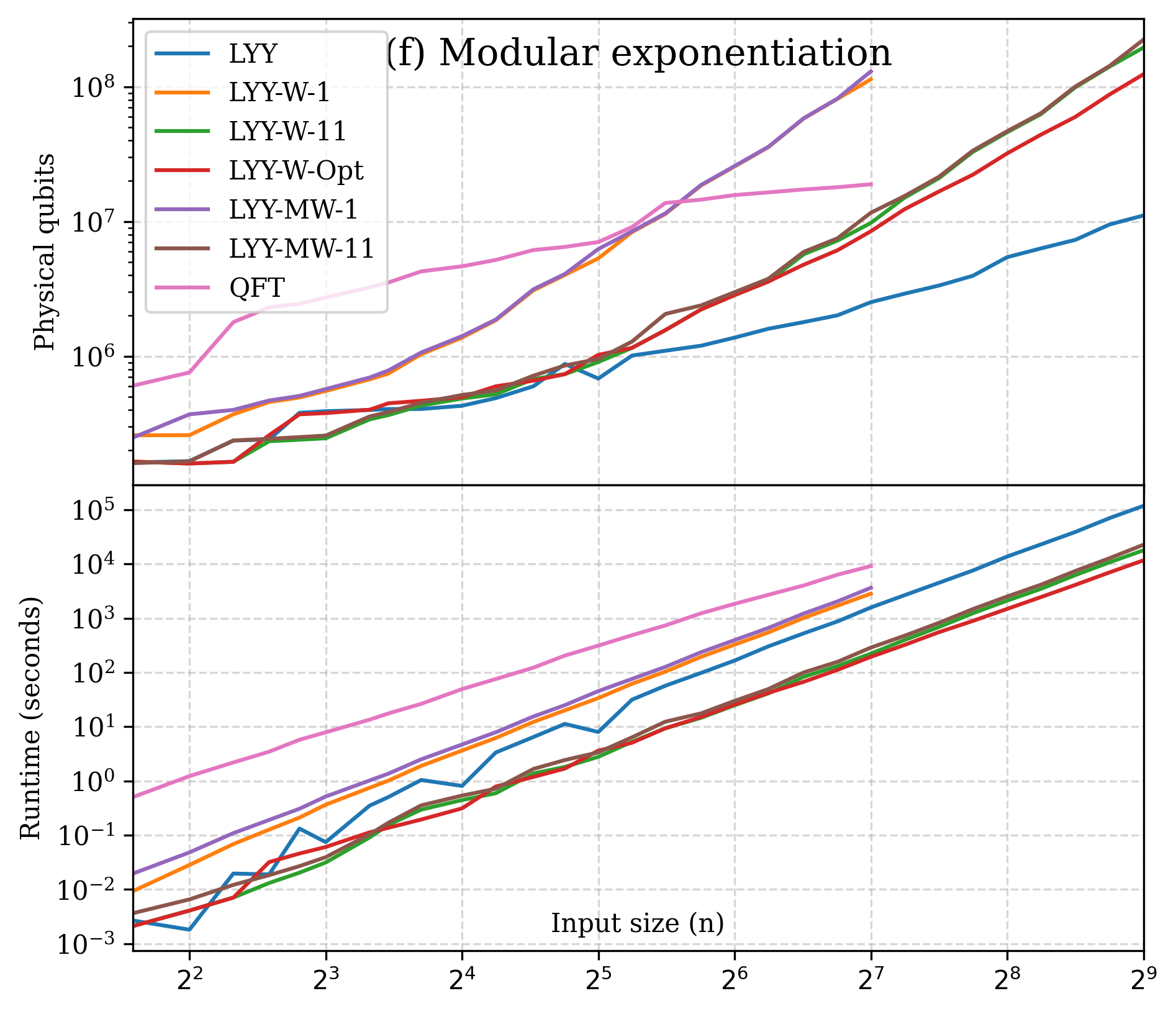}
    }
    \caption{Resource estimation for various quantum arithmetic algorithms}
    \label{fig:results}
\end{figure*}

\begin{figure*}
    \centering
    \begin{subfigure}[t]{0.48\textwidth}
        \centering
        \includegraphics[width=1.02\textwidth]{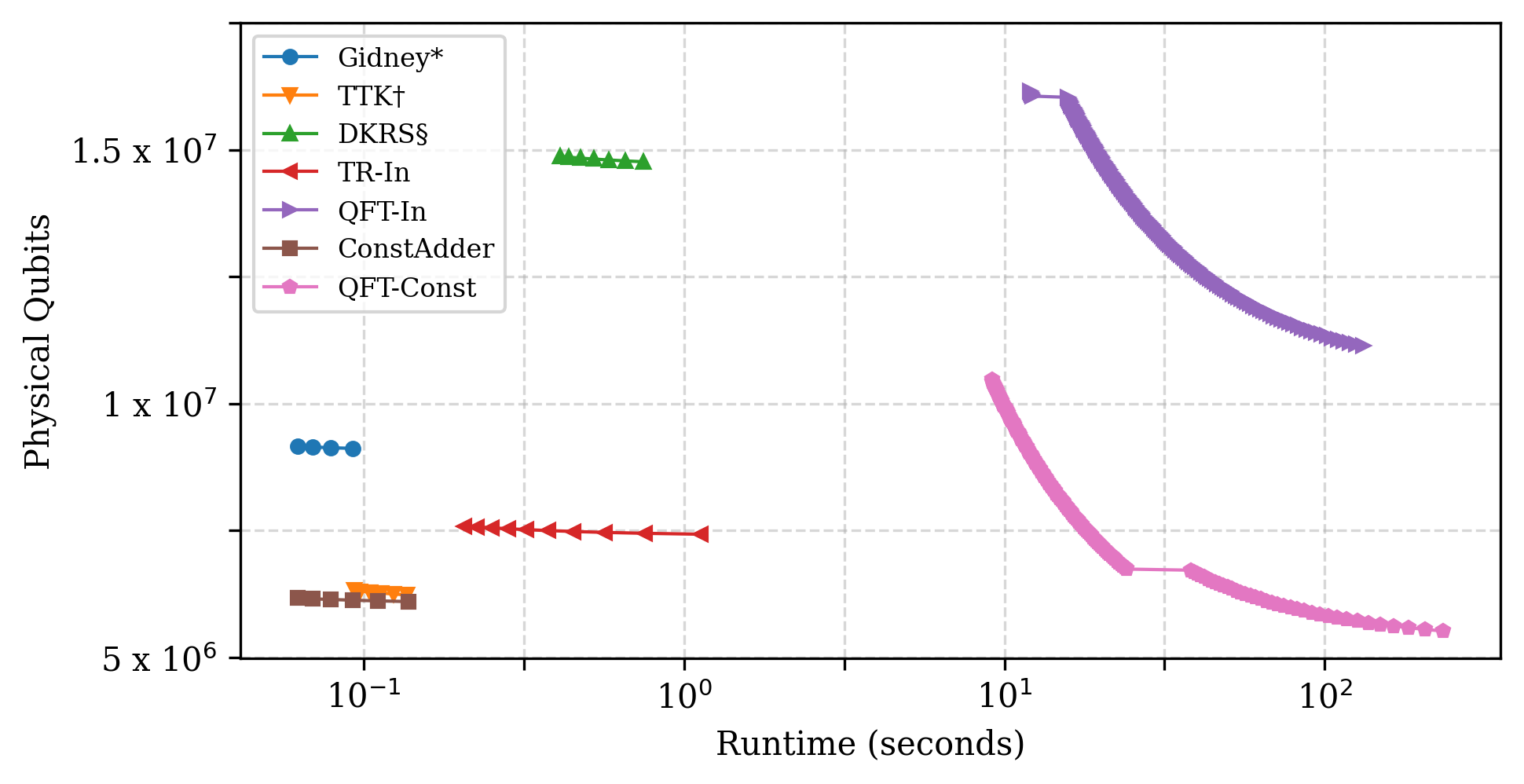}
        \caption{Adders with input size $n=2^{11}$.\\[0.8ex]       
        }
        \label{fig:pareto_adder}
    \end{subfigure}
    \hfill
    \begin{subfigure}[t]{0.48\textwidth} 
        \centering
        \includegraphics[width=1.02\textwidth]{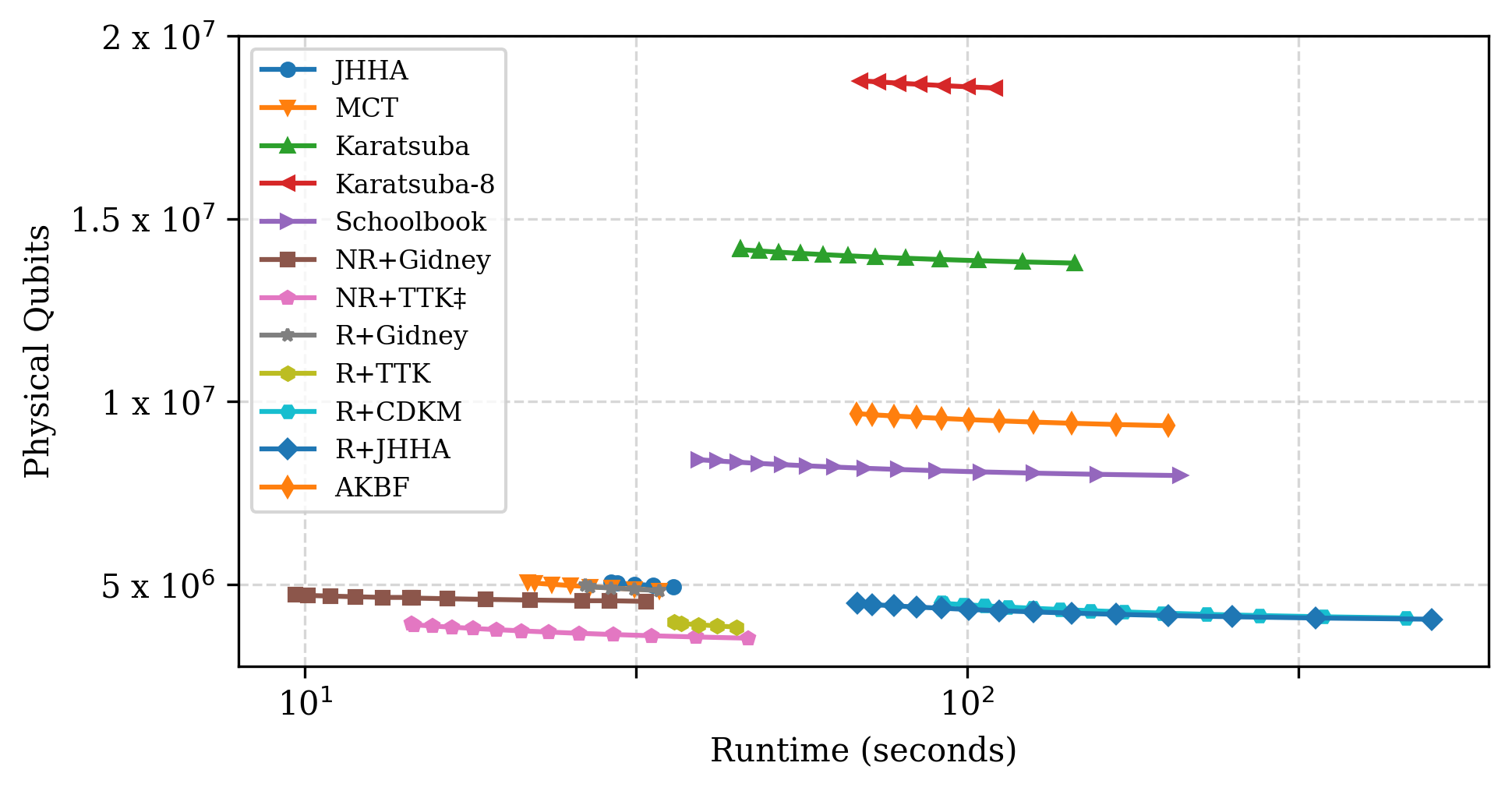}
        \caption{Multipliers and dividers with input size $n=2^9$\\[0.8ex]
        }
        \label{fig:pareto_mult}
    \end{subfigure}
    \caption{Pareto frontier estimations for various quantum arithmetic algorithms}
    \label{fig:pareto}
\end{figure*}

\section{Results and Discussion} \label{sec:ResultsDiscussion}

In this paper, we implemented six classes of quantum arithmetic algorithms and identified the best-performing ones in terms of runtime, qubit usage, and space-time trade-offs (\S \ref{cost_analysis}). We also optimized their performance, using the parameter fine-tuning of the windowed ModExp algorithm and the adder component space exploration in division as examples. This work demonstrates multiple applications of resource estimation for studying and improving quantum algorithms, with examples detailed in this section:
\begin{itemize}
  \item Choosing the best algorithm for a given task (\S \ref{cost_analysis}).
  \item Exploring space-time trade-offs (\S \ref{time_space_trade_offs}).
  \item Selecting optimal components (\S \ref{design_space_exploration}).  
  \item Fine-tuning algorithm parameters (\S \ref{parameter_tuning}).
  \item Identifying ``tipping points" where one algorithm outperforms another (\S \ref{tipping_point_analysis}).
  \item Analyzing asymptotic complexity (\S \ref{asymptotic_complexity_analysis}).
\end{itemize}

In addition, the resource estimator allows researchers to compare qubit technologies, error correction schemes, and other architectural features \cite{Basov_2023}, but we choose to focus on algorithmic applications in this paper.

\subsection{Cost Analysis and Comparison of Quantum Arithmetic Algorithms}
\label{cost_analysis}

Figure \ref{fig:results} highlights the estimates for two key metrics: number of physical qubits and runtime. Based on these metrics, we identified the best-performing algorithms in each class.

\begin{itemize}
\item \textbf{In-place addition.} In figure \ref{fig:results}a, TTK \cite{takahashi2009quantum}, CDKM  \cite{cuccaro2004new}, JHHA \cite{jayashree2016ancilla} and TR \cite{thapliyal2013design} adders are the best by physical qubits count, with zero or very small difference between them. Gidney adder  \cite{gidney2018halving} is fastest by runtime.
\item \textbf{Out-of-place addition.} Figure \ref{fig:results}b shows how Gidney \cite{gidney2018halving}, Gayathri \cite{gayathri2021} and Wang \cite{wang2016improved} adders lead both in physical qubit count and runtime, with identical performance. The HiRadix \cite{wang2023higher} adder demonstrates an advantage in T-depth due to parallelization, but it is not reflected in runtime estimates.
\item \textbf{Quantum-classical addition.} As shown in figure \ref{fig:results}c, ConstAdder \cite{fedoriaka2025} outperforms in both physical qubit count and runtime, though Gidney adder is very close with it in runtime. Notably, the QFT adder surpasses others in physical qubit count starting from $n \approx 7000$.
\item \textbf{Multiplication.} In figure \ref{fig:results}d, for small $n$, the MCT multiplier \cite{munozcoreas2017} is fastest, and both JHHA \cite{jayashree2016ancilla} and MCT \cite{munozcoreas2017} multipliers perform best in physical qubit count. Starting from $n=2^{12}$, the Karatsuba multiplier \cite{gidney2019karatsuba}, with sub-quadratic theoretical complexity, dominates in runtime as expected. However, its efficiency decreases for non-power-of-2 values of $n$ due to the need for padding qubits to round $n$ up to the next power of 2. The extrapolated trend suggests that Karatsuba algorithm's runtime will consistently outperform all others for $n \geq 2^{18}$.
\item \textbf{Division.} In figure \ref{fig:results}e, non-restoring divider using TTK adder \cite{takahashi2009quantum} is the best by physical qubit count. Non-restoring divider using Gidney adder is the best by runtime, mirroring the trend observed in in-place addition. Compared to their restoring division counterparts, non-restoring dividers require fewer physical qubits and achieve faster runtimes. This is because restoring division involves additional steps to revert the result after a failed subtraction. The performance of both algorithms is strongly influenced by the choice of adder primitives.
\item \textbf{Modular exponentiation.} In figure \ref{fig:results}f, LYY algorithm \cite{liu2021cnot} is the best by physical qubits count. LYY-W-Opt algorithm is the best by runtime.
\end{itemize}

\subsection{Space-Time Trade-Offs with Pareto Frontiers}
\label{time_space_trade_offs}
Figure \ref{fig:pareto} illustrates space-time diagrams from Pareto frontier estimations, where each point represents an optimal qubit count and runtime pair for a circuit. These diagrams help identify configurations that meet hardware constraints and compare space-time trade-offs across algorithms.

Figure \ref{fig:pareto_adder} shows that QFT-based algorithms tend to be slower and more resource-intensive, but offer greater space-time trade-off by allowing significant qubit count reductions at the cost of much longer runtimes. This is due to the QFT-based algorithms using rotations, which require more T gates. This more pronounced trade-off is directly seen in figure \ref{fig:pareto_adder} as the quantum-classical ConstAdder achieves the lowest physical qubit count and the fastest runtime simultaneously\cite{fedoriaka2025}, but the QFT-Const adder uses fewer qubits than the ConstAdder when its runtime is extended to over a minute \cite{draper2000}\cite{fedoriaka2025}.

Figure \ref{fig:pareto_mult} highlights the results of non-QFT based multipliers and dividers. The slopes of the non-QFT algorithms are very small, on the order of $10^{-4}$ to $10^{-5}$, revealing that even with increased runtime, there is only a minor change in the physical qubit requirements. Therefore, from a T-factory perspective the algorithms are already optimized.

\subsection{Design Space Exploration}
\label{design_space_exploration}
In quantum algorithms, design space exploration involves identifying all possible parameter configurations and primitive choices for implementing an algorithm, then evaluating their performance in terms of resource usage and runtime. For example, we explored various in-place adders as building blocks for dividers and found that the TTK adder minimizes physical qubit count, aligning with the previous theoretical results \cite{takahashi2009quantum}\cite{wang2024boosting}. This approach can be generalized to a multi-dimensional search, testing combinations of building blocks and their alternatives to identify optimal designs.

\subsection{Optimization of Algorithmic Parameters}
\label{parameter_tuning}
Some algorithms have tunable parameters that influence their operation, such as the radix size in higher radix adders and the window size in modular exponentiation. By exploring different values of these parameters, one can identify the configuration that yields the best performance. 

For example, in the ModExp algorithm, we found the optimal window size to minimize estimated runtime. By fixing $n=32$ and varying the window size $w$, we found that windowed ModExp using fast modular multiplication and $w=11$ (LYY-W-11) has the fastest runtime. However, the optimal value of $w$ varies as $n$ changes. The theoretical runtime of this algorithm is $$T(n,w) \approx (c_1 \cdot 2^w + c_2 \cdot n^2) \cdot \frac{n}{w}.$$ We used the resource estimator results to fit $c_1, c_2$ using the least squares method. Solving the condition $\frac{\partial T}{\partial w}=0$ suggested that the optimal window size is approximately $w \approx 2 \cdot \log_2(n)$. Numerical checks showed that this formula deviates from $\arg \underset{w}{\max}~ T(n,w)$ no more than by $2.08$. We designed the LYY-W-Opt algorithm with the best estimated runtime using window size: $$w = \lfloor 2 \log_2(n) + 0.5 \rfloor.$$

\subsection{Tipping Point Identification}
\label{tipping_point_analysis}
A tipping point occurs when one algorithm becomes more efficient or preferable than another as the problem size increases. For example, in multiplication algorithms, straightforward Shift-and-Add multiplication is faster than Karatsuba multiplication for $n < 2^{12}$. However, for $n \ge 2^{12}$, Karatsuba multiplication becomes more efficient, especially when $n$ is a power of 2. Therefore, $n=2^{12}$ marks the tipping point. 

\subsection{Asymptotic Complexity Analysis} 
\label{asymptotic_complexity_analysis}
Arithmetic algorithms typically have $O(n^a)$ complexity. Thus, plotting runtime on a log-log scale produces a linear relationship, where we can use the least squares method to find the slope $a$ and compare it with theoretical predictions. 

Adders show $a \approx 1.07 \dots 1.12$, which is consistent with $O(n)$ complexity. Shift-and-Add multipliers have $a \approx 2.10$, and dividers have $a \approx 2.10 \dots 2.12$, aligning with $O(n^2)$ complexity. For ModExp with optimal window size, $a \approx 2.97$, consistent with $O(n^3)$ complexity. Karatsuba multiplication gives $a \approx 1.76$, slightly slower than the theoretical complexity of classical Karatsuba multiplication $O(n^{log_2 3}) \approx O(n^{1.58})$ \cite{karatsuba1962multiplication}, but still sub-quadratic.


\subsection{Insights for Optimizing Arithmetic Algorithms}

This section summarizes lessons we learned about optimizing arithmetic algorithms for implementation and resource estimation:

\textbf{Parallel Qubit Operation.} While previous studies, such as HiRadix\cite{wang2023higher}, use T-depth estimation to demonstrate how parallelization reduces runtime, these effects are not reflected in the estimation results. Azure Quantum Resource Estimator uses a compilation scheme, Parallel Synthesis Sequential Pauli Computation (PSSPC), which manages the layout, routing, and scheduling of the algorithm onto the surface code\cite{beverland2022}. By default, a conservative approach is taken to provide an upper bound and does not take into account parallel structures in algorithms. Therefore, when implementing the quantum higher radix adder, we observed an increased qubit count for setting up the parallel structure, along with a slowdown compared to other adders.

\textbf{Reset Operation.} Reset operations, common in classical computing, destroy any quantum entanglement or superposition involving the target qubit, which is a critical limitation in quantum algorithm design. While re-implementing algorithms from studies like \cite{Yuan2022ANF}, we tested with inputs in superposition and confirmed that the reset operation disrupted entangled states and introduced errors.

\textbf{Uncomputation.} In reversible circuits, uncomputation is used to clean up temporary changes to ancilla qubits so that they can be re-used. Some studies, like \cite{TR2009Sub}, omitted details on uncomputation, leaving ancilla as unhandled garbage qubits. To ensure a fair comparison between circuits with and without proper uncomputation, we applied Bennet's garbage removal scheme \cite{Bennet}, which involves computing results on an ancilla, copying them using CNOT gates, and then uncomputing the ancilla. This modification leads to at least a twofold increase in runtime as the circuit is repeated in reverse. Unfortunately, the Bennet scheme is not feasible for HiRadix due to its hybrid CLA and RCA structure\cite{wang2023higher}, so the adder was modified from an in-place to an out-of-place design.

\section{Conclusion}

As Quantum Computing technology continues to advance, the demand for optimized arithmetic circuits is expected to grow.  In this work, we implement a comprehensive set of quantum arithmetic algorithms and estimate their resource requirements and runtime using the Azure Quantum Resource Estimator. We analyze the asymptotic behavior of various algorithms, optimize key components, and explore trade-offs between computation time and system scale. Our goal is to provide both a practical library and a valuable knowledge base for the broader quantum computing community to effectively incorporate the latest arithmetic algorithms into their applications. Additionally, we showcase applications of the resource estimator as a tool to advance quantum computing research.

\section*{Acknowledgments}

We sincerely thank Mariia Mykhailova for her mentorship. We also appreciate the Quantum Open Source Foundation for hosting the mentorship program that connects quantum enthusiasts with mentors from academia and industry.
\bibliographystyle{IEEEtran}
\bibliography{refs}

\begin{thebibliography}{10}
\providecommand{\url}[1]{#1}
\csname url@samestyle\endcsname
\providecommand{\newblock}{\relax}
\providecommand{\bibinfo}[2]{#2}
\providecommand{\BIBentrySTDinterwordspacing}{\spaceskip=0pt\relax}
\providecommand{\BIBentryALTinterwordstretchfactor}{4}
\providecommand{\BIBentryALTinterwordspacing}{\spaceskip=\fontdimen2\font plus
\BIBentryALTinterwordstretchfactor\fontdimen3\font minus \fontdimen4\font\relax}
\providecommand{\BIBforeignlanguage}[2]{{%
\expandafter\ifx\csname l@#1\endcsname\relax
\typeout{** WARNING: IEEEtran.bst: No hyphenation pattern has been}%
\typeout{** loaded for the language `#1'. Using the pattern for}%
\typeout{** the default language instead.}%
\else
\language=\csname l@#1\endcsname
\fi
#2}}
\providecommand{\BIBdecl}{\relax}
\BIBdecl

\bibitem{beverland2022}
\BIBentryALTinterwordspacing
M.~E. Beverland, P.~Murali, M.~Troyer, K.~M. Svore, T.~Hoefler, V.~Kliuchnikov, G.~H. Low, M.~Soeken, A.~Sundaram, and A.~Vaschillo, ``Assessing requirements to scale to practical quantum advantage,'' 2022. [Online]. Available: \url{https://arxiv.org/abs/2211.07629}
\BIBentrySTDinterwordspacing

\bibitem{hansen2023}
\BIBentryALTinterwordspacing
E.~Hansen, S.~Joshi, and H.~Rarick, ``Resource estimation of quantum multiplication algorithms,'' in \emph{2023 IEEE International Conference on Quantum Computing and Engineering (QCE)}.\hskip 1em plus 0.5em minus 0.4em\relax IEEE, Sep. 2023, p. 199–202. [Online]. Available: \url{http://dx.doi.org/10.1109/QCE57702.2023.10211}
\BIBentrySTDinterwordspacing

\bibitem{ArithmeticReview2024}
\BIBentryALTinterwordspacing
S.~Wang, X.~Li, W.~J.~B. Lee, S.~Deb, E.~Lim, and A.~Chattopadhyay, ``A comprehensive study of quantum arithmetic circuits,'' 2024. [Online]. Available: \url{https://arxiv.org/abs/2406.03867}
\BIBentrySTDinterwordspacing

\bibitem{Azure_Quantum_Resource_Estimator}
\BIBentryALTinterwordspacing
W.~van Dam, M.~Mykhailova, and M.~Soeken, ``{Using Azure Quantum Resource Estimator for Assessing Performance of Fault Tolerant Quantum Computation},'' in \emph{Proceedings of the SC '23 Workshops of The International Conference on High Performance Computing, Network, Storage, and Analysis}, ser. SC-W '23.\hskip 1em plus 0.5em minus 0.4em\relax New York, NY, USA: Association for Computing Machinery, 2023, p. 1414–1419. [Online]. Available: \url{https://doi.org/10.1145/3624062.3624211}
\BIBentrySTDinterwordspacing

\bibitem{cuccaro2004new}
\BIBentryALTinterwordspacing
S.~A. Cuccaro, T.~G. Draper, S.~A. Kutin, and D.~P. Moulton, ``A new quantum ripple-carry addition circuit,'' 2004. [Online]. Available: \url{https://arxiv.org/abs/quant-ph/0410184}
\BIBentrySTDinterwordspacing

\bibitem{gidney2018halving}
\BIBentryALTinterwordspacing
C.~Gidney, ``Halving the cost of quantum addition,'' \emph{Quantum}, vol.~2, p.~74, Jun. 2018. [Online]. Available: \url{http://dx.doi.org/10.22331/q-2018-06-18-74}
\BIBentrySTDinterwordspacing

\bibitem{takahashi2009quantum}
\BIBentryALTinterwordspacing
Y.~Takahashi, S.~Tani, and N.~Kunihiro, ``Quantum addition circuits and unbounded fan-out,'' 2009. [Online]. Available: \url{https://arxiv.org/abs/0910.2530}
\BIBentrySTDinterwordspacing

\bibitem{draper2006logarithmic}
\BIBentryALTinterwordspacing
T.~G. Draper, S.~A. Kutin, E.~M. Rains, and K.~M. Svore, ``A logarithmic-depth quantum carry-lookahead adder,'' \emph{Quantum Information and Computation}, vol.~6, no. 4\&5, 2006. [Online]. Available: \url{http://dx.doi.org/10.26421/QIC6.4-5-4}
\BIBentrySTDinterwordspacing

\bibitem{jayashree2016ancilla}
H.~Jayashree, H.~Thapliyal, H.~R. Arabnia, and V.~K. Agrawal, ``Ancilla-input and garbage-output optimized design of a reversible quantum integer multiplier,'' \emph{The Journal of Supercomputing}, vol.~72, pp. 1477--1493, 2016.

\bibitem{thapliyal2013design}
H.~Thapliyal and N.~Ranganathan, ``Design of efficient reversible logic-based binary and bcd adder circuits,'' \emph{ACM Journal on Emerging Technologies in Computing Systems (JETC)}, vol.~9, no.~3, pp. 1--31, 2013.

\bibitem{draper2000}
\BIBentryALTinterwordspacing
T.~G. Draper, ``Addition on a quantum computer,'' 2000. [Online]. Available: \url{https://arxiv.org/abs/quant-ph/0008033}
\BIBentrySTDinterwordspacing

\bibitem{qsharpStdLib}
\BIBentryALTinterwordspacing
``Q\# standard library.'' [Online]. Available: \url{https://github.com/microsoft/qsharp/tree/main/library/std}
\BIBentrySTDinterwordspacing

\bibitem{cheng2002quantum}
K.-W. Cheng and C.-C. Tseng, ``Quantum full adder and subtractor,'' \emph{Electronics Letters}, vol.~38, no.~22, pp. 1343--1344, 2002.

\bibitem{LingAdder2023}
S.~Wang and A.~Chattopadhyay, ``Reducing depth of quantum adder using {Ling} structure,'' in \emph{2023 IFIP/IEEE 31st International Conference on Very Large Scale Integration (VLSI-SoC)}, 2023, pp. 1--6.

\bibitem{gayathri2021}
S.~Gayathri, R.~Kumar, S.~Dhanalakshmi, B.~K. Kaushik, and M.~Haghparast, ``{T}-count optimized wallace tree integer multiplier for quantum computing,'' \emph{International Journal of Theoretical Physics}, vol.~60, no.~8, pp. 2823--2835, 2021.

\bibitem{wang2016improved}
F.~Wang, M.~Luo, H.~Li, Z.~Qu, and X.~Wang, ``Improved quantum ripple-carry addition circuit,'' \emph{Science China Information Sciences}, vol.~59, pp. 1--8, 2016.

\bibitem{wang2023higher}
S.~Wang, A.~Baksi, and A.~Chattopadhyay, ``A higher radix architecture for quantum carry-lookahead adder,'' \emph{Scientific Reports}, vol.~13, no.~1, p. 16338, 2023.

\bibitem{liu2021cnot}
\BIBentryALTinterwordspacing
X.~Liu, H.~Yang, and L.~Yang, ``{CNOT}-count optimized quantum circuit of the {Shor's} algorithm,'' 2021. [Online]. Available: \url{http://arxiv.org/abs/2112.11358}
\BIBentrySTDinterwordspacing

\bibitem{fedoriaka2025}
\BIBentryALTinterwordspacing
D.~Fedoriaka, ``New circuit for quantum adder by constant,'' 2025. [Online]. Available: \url{https://arxiv.org/abs/2501.07060}
\BIBentrySTDinterwordspacing

\bibitem{pavlidis2014fast}
\BIBentryALTinterwordspacing
A.~Pavlidis and D.~Gizopoulos, ``Fast quantum modular exponentiation architecture for {Shor's} factorization algorithm,'' \emph{Quantum Information and Computation}, vol.~14, no. 7\&8, 2014. [Online]. Available: \url{https://dl.acm.org/doi/10.5555/2638682.2638690}
\BIBentrySTDinterwordspacing

\bibitem{munozcoreas2017}
\BIBentryALTinterwordspacing
E.~Muñoz-Coreas and H.~Thapliyal, ``{T}-count optimized design of quantum integer multiplication,'' 2017. [Online]. Available: \url{https://arxiv.org/abs/1706.05113}
\BIBentrySTDinterwordspacing

\bibitem{gidneyKaratsubaRepo}
\BIBentryALTinterwordspacing
 [Online]. Available: \url{https://github.com/Strilanc/quantum-karatsuba-2019}
\BIBentrySTDinterwordspacing

\bibitem{gidney2019karatsuba}
\BIBentryALTinterwordspacing
C.~Gidney, ``Asymptotically efficient quantum {Karatsuba} multiplication,'' 2019. [Online]. Available: \url{https://arxiv.org/abs/1904.07356}
\BIBentrySTDinterwordspacing

\bibitem{orts2023improving}
F.~Orts, E.~Filatovas, G.~Ortega, J.~SanJuan-Estrada, and E.~Garz{\'o}n, ``Improving the number of {T} gates and their spread in integer multipliers on quantum computing,'' \emph{Physical Review A}, vol. 107, no.~4, p. 042621, 2023.

\bibitem{thapliyal2019quantum}
H.~Thapliyal, E.~Munoz-Coreas, T.~Varun, and T.~S. Humble, ``Quantum circuit designs of integer division optimizing {T}-count and {T}-depth,'' \emph{IEEE transactions on emerging topics in computing}, vol.~9, no.~2, pp. 1045--1056, 2019.

\bibitem{khosropour2011quantum}
\BIBentryALTinterwordspacing
A.~Khosropour, H.~Aghababa, and B.~Forouzandeh, ``Quantum division circuit based on restoring division algorithm,'' in \emph{Eighth International Conference on Information Technology: New Generations}, 2011. [Online]. Available: \url{http://dx.doi.org/10.1109/ITNG.2011.177}
\BIBentrySTDinterwordspacing

\bibitem{Thapliyal2016MappingOS}
H.~Thapliyal, ``Mapping of subtractor and adder-subtractor circuits on reversible quantum gates,'' \emph{Transactions on Computational Science XXVII}, vol.~27, pp. 10--34, 2016.

\bibitem{wang2024boosting}
\BIBentryALTinterwordspacing
S.~Wang, E.~Lim, and A.~Chattopadhyay, ``Boosting the efficiency of quantum divider through effective design space exploration,'' 2024. [Online]. Available: \url{https://arxiv.org/abs/2403.01206}
\BIBentrySTDinterwordspacing

\bibitem{shor1994algorithms}
P.~W. Shor, ``Algorithms for quantum computation: discrete logarithms and factoring,'' in \emph{Proceedings 35th annual symposium on foundations of computer science}.\hskip 1em plus 0.5em minus 0.4em\relax Ieee, 1994, pp. 124--134.

\bibitem{li2014class}
X.~Li, G.~Yang, C.~M. Torres~Jr, D.~Zheng, and K.~L. Wang, ``A class of efficient quantum incrementer gates for quantum circuit synthesis,'' \emph{International Journal of Modern Physics B}, vol.~28, no.~01, p. 1350191, 2014.

\bibitem{TR2009Sub}
H.~Thapliyal and N.~Ranganathan, ``Design of efficient reversible binary subtractors based on a new reversible gate,'' in \emph{2009 IEEE Computer Society Annual Symposium on VLSI}, 2009, pp. 229--234.

\bibitem{Xia2019NovelMQ}
\BIBentryALTinterwordspacing
H.~Xia, H.~Li, H.~Zhang, Y.~Liang, and J.~Xin, ``Novel multi-bit quantum comparators and their application in image binarization,'' \emph{Quantum Information Processing}, vol.~18, 2019. [Online]. Available: \url{https://api.semanticscholar.org/CorpusID:191136795}
\BIBentrySTDinterwordspacing

\bibitem{babbush2018encoding}
R.~Babbush, C.~Gidney, D.~W. Berry, N.~Wiebe, J.~McClean, A.~Paler, A.~Fowler, and H.~Neven, ``Encoding electronic spectra in quantum circuits with linear {T} complexity,'' \emph{Physical Review X}, vol.~8, no.~4, p. 041015, 2018.

\bibitem{munoz2018squareRoot}
E.~Mu{\~n}oz-Coreas and H.~Thapliyal, ``{T}-count and qubit optimized quantum circuit design of the non-restoring square root algorithm,'' \emph{ACM Journal on Emerging Technologies in Computing Systems (JETC)}, vol.~14, no.~3, pp. 1--15, 2018.

\bibitem{saeedi2013}
\BIBentryALTinterwordspacing
M.~Saeedi and I.~L. Markov, ``Quantum circuits for {GCD} computation with {$O(n \log n)$} depth and {$O(n)$} ancillae,'' 2013. [Online]. Available: \url{https://arxiv.org/abs/1304.7516}
\BIBentrySTDinterwordspacing

\bibitem{Svore_2018}
\BIBentryALTinterwordspacing
K.~Svore, A.~Geller, M.~Troyer, J.~Azariah, C.~Granade, B.~Heim, V.~Kliuchnikov, M.~Mykhailova, A.~Paz, and M.~Roetteler, ``{Q\#: Enabling Scalable Quantum Computing and Development with a High-level DSL},'' in \emph{Proceedings of the Real World Domain Specific Languages Workshop 2018}, ser. RWDSL2018.\hskip 1em plus 0.5em minus 0.4em\relax ACM, Feb. 2018. [Online]. Available: \url{http://dx.doi.org/10.1145/3183895.3183901}
\BIBentrySTDinterwordspacing

\bibitem{Lopez_2024}
\BIBentryALTinterwordspacing
S.~Lopez, M.~Mykhailova, and B.~Benefield, ``Resource estimator target parameters - {A}zure {Q}uantum.'' [Online]. Available: \url{https://learn.microsoft.com/en-us/azure/quantum/overview-resources-estimator}
\BIBentrySTDinterwordspacing

\bibitem{Basov_2023}
\BIBentryALTinterwordspacing
I.~Basov, ``Modeling quantum architecture with {A}zure {Q}uantum resource estimator,'' Nov 2023. [Online]. Available: \url{https://devblogs.microsoft.com/qsharp/modeling-quantum-architecture-with-azure-quantum-resource-estimator/}
\BIBentrySTDinterwordspacing

\bibitem{karatsuba1962multiplication}
A.~A. Karatsuba and Y.~P. Ofman, ``Multiplication of many-digital numbers by automatic computers,'' in \emph{Doklady Akademii Nauk}, vol. 145, no.~2.\hskip 1em plus 0.5em minus 0.4em\relax Russian Academy of Sciences, 1962, pp. 293--294.

\bibitem{Yuan2022ANF}
\BIBentryALTinterwordspacing
S.~Yuan, S.~Gao, C.~Wen, Y.~Wang, H.~Qu, and Y.~Wang, ``A novel fault-tolerant quantum divider and its simulation,'' \emph{Quantum Information Processing}, vol.~21, 2022. [Online]. Available: \url{https://api.semanticscholar.org/CorpusID:248877336}
\BIBentrySTDinterwordspacing

\bibitem{Bennet}
C.~H. Bennett, ``Logical reversibility of computation,'' \emph{IBM Journal of Research and Development}, vol.~17, no.~6, pp. 525--532, 1973.

\end{thebibliography}

\end{document}